\documentclass[preprint]{ptephy_v1}
\usepackage{amssymb}
\usepackage{graphicx}
\numberwithin{equation}{section}

\newcommand{\cb}{\bar{c}}
\newcommand{\lb}[1]{\label{#1}}
\newcommand{\cl}{\mathcal{L}}
\newcommand{\bb}{\bar{B}}

\newcommand{\vp}{\varphi}
\newcommand{\ptl}{\partial}

\def\vec#1{\mbox{\boldmath $#1$}}
\newcommand{\ct}{\tilde{C}}
\newcommand{\bt}{\tilde{B}}
\newcommand{\ch}{\mathcal{H}}
\newcommand{\cvp}{\tilde{\varphi}}
\newcommand{\cc}{\mathcal{C}}
\newcommand{\mcb}{\mathcal{B}}
\newcommand{\hh}{\mathcal{H}}

\begin{document}

\title{Massive dual gauge field and confinement in Minkowski space : Electric charge
}


\author{\name{\fname{Hirohumi} \surname{Sawayanagi}}{1}}

\address{\affil{1}{National Institute of Technology, Kushiro College, Kushiro, 084-0916, Japan}
\email{sawa@kushiro-ct.ac.jp}}


\begin{abstract}
     SU(2) gauge theory in the nonlinear gauge of the Curci-Ferrari type is studied in low-energy region.  
We give a classical solution that connects color electric charges.  Its dual solution, 
which has a configuration of monopole, is also presented.  
Due to the gluon condensation subsequent to the ghost condensation, 
these classical fields become massive.  
The massive Lagrangian with the classical solution and that with the dual solution are 
derived.  We show that these Lagrangians produce a linear potential between a quark and an antiquark.  
This is the mechanism of quark confinement that is different from the 
magnetic monopole condensation.  
\end{abstract}

\subjectindex{B0,B3,B6}

\maketitle

\section{Introduction}

     In the dual superconductor picture of quark confinement, magnetic monopoles are 
necessary (see, e.g., \cite{rip}).  Just like a Cooper pair in superconductivity, they must condense in the vacuum.  
This condensation gives a mass for 
non-Abelian gauge fields, and a linear potential between a quark and an antiquark is expected.  

     In Ref.~\cite{hs}, we studied the SU(2) gauge theory in the nonlinear gauge of the Curci-Ferrari type, and 
proposed another mechanism that gives a mass for gauge fields.  
In the low-energy region 
below the QCD scale parameter $\Lambda_{\mathrm{QCD}}$, the ghost condensation happens, and 
the SU(2) gauge theory breaks down to the U(1) theory \cite{hs1}.  If we choose the unbroken 
U(1) in the $A=3$ direction in SU(2), an additional condensate $\langle A_{\mu}^+A^{-\mu}\rangle$ appears.  
Because of this condensate, although the quantum U(1) gauge field $a_{\mu}^3$ 
is massless, the classical part $b_{\mu}^3$ acquires the mass $m$.  

     In the previous paper \cite{hs2}, we considered the magnetic potential $\ct_{\mu}$ as the classical part $b_{\mu}^3$.  
It was shown that the color magnetic charges $Q_m$ and $-Q_m$ are confined by the linear potential.  
We also introduced the dual magnetic potential $\cc_{\mu}$ consistently, 
and derived the same linear potential.   

     In this paper, based on Ref.~\cite{hs2}, we study the confinement of the color electric charges $Q_e$ and $-Q_e$.  
In the next section, we briefly review Ref.~\cite{hs2}.  In Sect.~3, we introduce the classical gauge 
field $\bt_{\mu}$ which couples with the color electric current $j^{\mu}$.  We call it the electric potential.  
Its dual potential $\mcb_{\mu}$ is also defined.  
Referring to $\ct_{\mu}$ and $\cc_{\mu}$, we present the relation between $\bt_{\mu}$ and $\mcb_{\mu}$.  
From this relation, the Lagrangian for $\bt_{\mu}$ and that for $\mcb_{\mu}$ are given.  
Using these Lagrangians, the linear potential between $Q_e$ and $-Q_e$ is derived in Sect.~4.  
The origin of the linear potential is discussed in Sect.~5, and the configuration which 
yields the quark confinement is discussed in Sect.~6.  In Sect.~7, the present theory is compared with 
the dual Ginzburg-Landau model of dual superconductor.  
Section~8 is devoted to a summary and comments.  
In Appendix~A, notations and formulas are summarized.  For a static magnetic charge, the solution of 
the equation of motion and its dual solution are presented in Appendix~B.  
The solution and its dual solution for a static electric charge are also given.  In Appendix~C, we calculate 
the integral which gives the linear potential.  To make the manuscript self-contained, the 
existence of the ghost condensation in Minkowski space is explained in Appendix~D.

\section{Magnetic potential and its dual potential}

     We review Ref.~\cite{hs2} briefly.  
Let us consider the SU(2) gauge theory with structure constants $f^{ABC}$.  
Using the notations 
\begin{align*}
 & F\cdot G=F^AG^A, \quad (F\times )^{AB}=f^{ACB}F^C,  \quad (F\times G)^A=f^{ABC}F^BG^C, \\
 & (\ptl\wedge A^A)_{\mu\nu}=\ptl_{\mu}A^A_{\nu}-\ptl_{\nu}A^A_{\mu}, \quad 
A=1,2,3,
\end{align*}
the Lagrangian is 
\[
     \cl_{\mathrm{inv}}(A)=-\frac{1}{4}G_{\mu\nu}\cdot G^{\mu\nu},\quad 
     G_{\mu\nu}^A=(\ptl\wedge A^A)_{\mu\nu}+g(A_{\mu}\times A_{\nu})^A.  
\]
This Lagrangian requires gauge fixing, and an appropriate gauge-fixing term and a ghost term are necessary.  
The Lagrangian for these terms is written as $\cl_{\varphi}(A)$.  

\subsection{SU(2) gauge theory in the low-energy region}

     In Refs.~\cite{hs1,hs3}, we employed the nonlinear gauge of the Curci-Ferrari type \cite{bt}.  Using 
the Nakanishi-Lautrup field $B^A$, 
the ghost $c^A$ and the antighost $\cb^A$, and the gauge parameter $\alpha_2$, we introduced 
the field $\vp^A=\alpha_2 (-B+ig\cb \times c)^A$.  
At the one-loop level, it was shown that $\vp^A$ acquires the vacuum expectation value (VEV) 
$\vp_0=|\langle \vp^A\rangle |\neq 0$ below the scale $\Lambda_{\mathrm{QCD}}$ \cite{hs1}.  
This phenomenon is called the ghost condensation \cite{hs3,sch,ks}.  
\footnote{In Minkowski space, although we could not show $\vp_0 \neq 0$ in 
Ref.~\cite{hs3}, we did it in Ref.~\cite{hs2}.  The treatments in these two articles 
are compared to stress $\vp_0 \neq 0$ in Appendix~D.  }
Choosing the VEV in the $A=3$ direction, we write $\langle \vp^A\rangle =\vp_0 \delta^{A3}$.  
Next we divided the gauge field $A_{\mu}^A$ into the classical part $b_{\mu}^A$ 
and the quantum part $a_{\mu}^A$ as 
\[ A_{\mu}^A= a_{\mu}^A + b_{\mu}^A, \quad b_{\mu}^A=b_{\mu}^3\delta^{A3}.  \]
In the presence of the VEV $\vp_0 \delta^{A3}$, 
ghost loops yield the tachyonic gluon masses for $a_{\mu}^A$ \cite{hs3, dv}.  
In Ref.~\cite{hs}, we have shown that the VEV $\langle A_{\mu}^+A^{-\mu}\rangle$ appears and 
the tachyonic gluon masses are removed.  Thus we obtained the Lagrangian 
$\cl(b+a)=\cl_{\mathrm{inv}}(b+a) + \cl_{\varphi}(a,b)$ with
\begin{align}
   \cl_{\mathrm{inv}}= &- \frac{1}{4}(F+H)^2+\frac{m^2}{2}[2a_{\mu}^3 b^{3\mu} + b_{\mu}^3 b^{3\mu} ] 
  + M^2a_{\mu}^+a^{-\mu} - \frac{g}{2}(F_{\mu\nu}+H_{\mu\nu})(a^{\mu}\times a^{\nu})^3 \nonumber \\
 &-\frac{g^2}{4}(a_{\mu}\times a_{\nu})^3(a^{\mu}\times a^{\nu})^3
     -\frac{1}{4}(\hat{D}_{\mu}a_{\nu}-\hat{D}_{\nu}a_{\mu})^a(\hat{D}^{\mu}a^{\nu}-\hat{D}^{\nu}a^{\mu})^a.  \lb{201}
\end{align}
Here, we used the notations $F_{\mu\nu}=(\ptl \wedge a^{3})_{\mu\nu}$, 
$H_{\mu\nu}=(\ptl \wedge b^{3})_{\mu\nu}$, $F^2=F_{\mu\nu}F^{\mu\nu}$, and 
$(\hat{D}_{\mu}a_{\nu})^a=(\ptl_{\mu}a_{\nu}+gA_{\mu}^3\times a_{\nu})^a$ $(a=1,2)$.  
We find, although the quantum part $a_{\mu}^3$ is massless, the classical part $b_{\mu}^3$ acquires the mass 
$m=\sqrt{g^3\vp_0/(32\pi)}$.  
At the one-loop level, the quantum parts $a_{\mu}^{a}\ (a=1,2)$ also acquire the mass $M$ defined by 
\[
     -\frac{m^2}{2g^2}= i\langle x|\mathrm{tr}\left(\Delta+M^2 \right)^{-1}|x \rangle.  
\]
The gauge-fixing and ghost part becomes 
\begin{align*}
 \cl_{\vp}(a,b)=& \frac{\alpha_1}{2}B\cdot B 
 +B\cdot [D_{\mu}(b)a^{\mu}+\cvp]  \nonumber \\
 &+ i\cb\cdot [D_{\mu}(b)D^{\mu}(b+a) +g\vp_0\times  +g\cvp \times ]c 
- \frac{(\vp_0+\cvp)\cdot(\vp_0+\cvp)}{2\alpha_2},  
\end{align*}
where $\cvp^A=\vp^A-\vp_0\delta^{A3}$ is the quantum fluctuation, and $\alpha_1$ is another gauge parameter. 

     We note, if $\vp^A$ is integrated out, $\cl_{\vp}(a,b)$ gives 
\[
 \cl_{\mathrm{NL}} =  B\cdot D_{\mu}(b)a^{\mu} + i\cb\cdot[D_{\mu}(b)D^{\mu}(b+a)c]
 + \frac{\alpha_1}{2}B\cdot B + \frac{\alpha_2}{2}\bb\cdot \bb -B\cdot \langle \vp\rangle.   
\]
When the classical field $b_{\mu}^A=0$, $\cl_{\mathrm{NL}}$ represents the nonlinear gauge of the Curci-Ferrari 
type \cite{bt}, and the last term is required to keep the BRS symmetry in the presence of $\vp_0$ \cite{hs}.

\subsection{Lagrangian with the magnetic potential $\ct_{\mu}$}

     First, we consider the magnetic potential $\ct_{\mu}$, and set $b_{\mu}^3= \ct_{\mu}$.  This field satisfies the equation of motion
\begin{equation}
   \ptl^{\nu}H_{\mu\nu}=\epsilon_{\mu\nu\alpha\beta}\ptl^{\nu}\frac{n^{\alpha}}{n^{\rho} \ptl_{\rho}}k^{\beta}, \quad 
   H_{\mu\nu}=(\ptl\wedge \ct)_{\mu\nu},   \lb{202}
\end{equation}
where $k^{\beta}$ is the magnetic current, and the space-like vector $n^{\alpha}$ satisfies $n^{\alpha}n_{\alpha}=-1$.  
When the mass term for $\ct_{\mu}$ exists, the equation of motion changes from Eq.~(\ref{202}) to 
\begin{equation}
     (D_{m}^{-1})_{\mu\nu}\ct^{\nu}-
\epsilon_{\nu\alpha\mu\beta}\frac{n^{\alpha}\ptl^{\mu}}{n^{\rho}\ptl_{\rho}}k^{\beta}= 0, \quad 
     (D_{m}^{-1})_{\mu\nu}= g_{\mu\nu}(\square + m^2)-\ptl_{\mu}\ptl_{\nu}.  \lb{203}
\end{equation}
As an example, $n^{\alpha}$ and $k^{\nu}$ for the Dirac monopole are presented in Appendix~B.  
The solutions $\ct_{\mu}$ for Eqs.~(\ref{202}) and  (\ref{203}) are also given in this appendix.  

     To incorporate the current $k^{\nu}$ in the 
Lagrangian, we replace $(\ptl\wedge \ct)^{\mu\nu}$ with 
\footnote{Equation~(\ref{204}) is Zwanziger's field strength $F=(\ptl \wedge A)-(n\cdot \ptl)^{-1}(n\wedge j_g)^d$ 
in Ref.~\cite{zwa}.}
\begin{equation}
     (\ptl\wedge \ct)^{\mu\nu} -\epsilon^{\mu\nu\alpha\beta}\frac{n_{\alpha}}{n^{\rho}\ptl_{\rho}}k_{\beta}.  \lb{204}
\end{equation}
Then, performing this replacement and neglecting the components $a_{\mu}^{a} (a=1,2)$, 
Eq.~(\ref{201}) leads to the Abelian Lagrangian 
\begin{equation}
     \cl_{\mathrm{mAbel}}=-\frac{1}{4}F^2 
     -\frac{1}{4}\left(\ptl\wedge \ct -\epsilon^{\mu\nu\alpha\beta}\frac{n_{\alpha}}{n^{\rho}\ptl_{\rho}}k_{\beta}\right)^2 
     + \frac{m^2}{2}\ct_{\mu}\ct^{\mu}, \lb{205}
\end{equation}
where, because of the equation of motion (\ref{203}), the linear term of $a_{\mu}^3$ vanishes.  

     Now we neglect $a_{\mu}^3$.  In Ref.~\cite{hs2}, it is shown that Eq.~(\ref{205}) gives the magnetic current-current 
correlation 
\begin{equation}
 \cl_{kk}=
 -\frac{1}{2}k_{\mu}\frac{1}{\square + m^2}k^{\mu} 
   -\frac{1}{2}k^{\mu}\frac{m^2}{\square + m^2}\frac{n_{\alpha}n^{\alpha}}{(n^{\rho}\ptl_{\rho})^2}
\left(g_{\mu\nu}-\frac{n_{\mu}n_{\nu}}{n_{\sigma}n^{\sigma}}\right)k^{\nu}.   \lb{206}
\end{equation}
We choose the current 
\begin{equation}
     k^{\mu}(x)=Q_mg^{\mu 0}\{\delta(\vec{x}-\vec{a})-\delta(\vec{x}-\vec{b})\}, \lb{207}
\end{equation}
where the magnetic charge is $Q_m$, and the position of the static magnetic monopole (antimonopole) is $\vec{a}$ ($\vec{b}$).  
We write $\vec{r}=\vec{a}-\vec{b}$, $r=|\vec{r}|$ and $n^{\mu}=(0, \vec{n})$, and follow the procedure in Refs.~\cite{suz, mts, sst, sst2}.  
Then, when $\vec{r} \parallel \vec{n}$, 
the correlation (\ref{206}) gives the magnetic monopole-antimonopole potential 
\begin{align}
  V_{\mathrm{m}}(r) =& V_{\mathrm{mY}}(r)+V_{\mathrm{mL}}(r), \quad  V_{\mathrm{mY}}(r)= \frac{-Q_m^2}{4\pi}\frac{e^{-mr}}{r},\nonumber \\
  & 
  V_{\mathrm{mL}}(r)=\sigma_m r + O(e^{-mr}), \quad \sigma_m=\frac{Q_m^2m^2}{8\pi}\ln \left(\frac{m^2+m_{\chi}^2}{m^2}\right), 
  \lb{208}
\end{align}
where $m_{\chi}$ is the ultraviolet cutoff for the momentum components $\vec{q}_{T}$ that is perpendicular to $\vec{n}$.
\footnote{The scale $m_{\chi}$ comes from the energy that the ghost condensation disappears.}  
Thus the magnetic monopoles are confined by the linear potential $V_{\mathrm{mL}}(r)$.  

     The derivation of the linear potential will be discussed in Sect.~4.  

\subsection{Lagrangian with the dual magnetic potential $\cc_{\mu}$}

     If we consider the magnetic monopole, $\ct_{\mu}$ is the space-like potential.  So, Cho introduced a time-like potential,  
which is called the dual magnetic potential $\cc_{\mu}$ \cite{cho}.  
We define a dual field strength by 
\begin{equation}
     \ch^{\mu\nu}=\frac{1}{2}\epsilon^{\mu\nu\alpha\beta}H_{\alpha\beta}=\epsilon^{\mu\nu\alpha\beta}\ptl_{\alpha}\ct_{\beta}.  \lb{209}
\end{equation}
and give the dual magnetic potential $\cc_{\mu}$ by the relation \cite{hs2} 
\begin{align}
 \ch^{\mu\nu}&=\epsilon^{\mu\nu\alpha\beta}\ptl_{\alpha}\ct_{\beta}= (\ptl\wedge \cc)^{\mu\nu} + \Lambda_m^{\mu\nu},  \lb{210}\\
 \Lambda_{m}^{\mu\nu}&= -\frac{n^{\mu}}{n^{\rho}\ptl_{\rho}}\ptl_{\sigma}(\ptl\wedge \cc)^{\sigma\nu}
 +\frac{n^{\nu}}{n^{\rho}\ptl_{\rho}}\ptl_{\sigma}(\ptl\wedge \cc)^{\sigma\mu}.  \nonumber
\end{align}
As we show in Appendix~B, in the case of the Dirac monopole, $\Lambda_{m}^{\mu\nu}$ represents the Dirac string.  
Since Eq.~(\ref{210}) is invariant under the transformations 
$\ct_{\mu} \to \ct_{\mu}+ \ptl_{\mu} \varepsilon$ and $\cc_{\mu} \to \cc_{\mu}+ \ptl_{\mu} \vartheta$, 
we choose the gauges
\[
     n_{\mu}\ct^{\mu}=0, \quad  n_{\mu}\cc^{\mu}=0.  
\]
Then Eq.~(\ref{210}) leads to 
\begin{equation}
     \ct^{\nu}=\epsilon^{\nu\mu\alpha\beta}\frac{n_{\mu}\ptl_{\alpha}}{n^{\rho}\ptl_{\rho}}\cc_{\beta}.  \lb{211}
\end{equation}
Using Eqs.~(\ref{210}) and (\ref{211}), we can rewrite the Lagrangian (\ref{201}).  
After neglecting the components $a_{\mu}^{a} (a=1,2)$, we obtain the Abelian part \cite{hs2}
\begin{align}
   \cl'_{\mathrm{mAbel}}=&-\frac{1}{4}F^2
-\frac{1}{4}(\ptl \wedge \cc)^2 +\frac{m^2}{2} \cc_{\mu}\cc^{\mu} - \cc_{\mu}k^{\mu} +\Omega_k,  \lb{212} \\
   \Omega_k=& \frac{m^2}{2}\cc^{\mu}N_{\mu\nu}\left[\ptl_{\lambda}(\ptl \wedge \cc)^{\lambda\nu}
     +m^2\cc^{\nu}-2k^{\nu}\right],  \nonumber 
\end{align}
where 
\[
N_{\mu\nu}=\frac{n^{\alpha}n_{\alpha}}{(n^{\rho}\ptl_{\rho})^2}
\left(g_{\mu\nu}-\frac{n_{\mu}n_{\nu}}{n^{\beta}n_{\beta}}\right).  
\]
Even if $\cl'_{\mathrm{mAbel}}$ contains $\Omega_k$, the dual field $\cc_{\mu}$ satisfies the equation of motion 
\begin{equation}
     (D_{m}^{-1})_{\mu\nu}\cc^{\nu} -k^{\mu}=0.  \lb{213}
\end{equation}
We solve Eq.~(\ref{213}) as 
\begin{equation}
     \cc^{\mu}= D_m^{\mu\nu}k_{\nu}, \quad 
     D_m^{\mu\nu}=\frac{g^{\mu\nu}}{\square + m^2} + \frac{\ptl^{\mu}\ptl^{\nu}}{m^2(\square+m^2)},  \lb{214}
\end{equation}
and substitute Eq.~(\ref{214}) into Eq.~(\ref{212}).  After neglecting the quantum part $-F^2/4$, Eq.~(\ref{212}) 
also gives the correlation (\ref{206}).  Namely, based on the dual magnetic potential $\cc_{\mu}$, the 
same confining potential (\ref{208}) is obtained.

\section{Electric potential and its dual potential}

     Let us consider the color electric current $j^A_{\mu}=j_{\mu}\delta^{A3}$, which usually 
couples with the gauge field as $-A^A_{\mu}j^{A\mu}=-(a_{\mu}^3+b_{\mu}^3)j^{\mu}$.  
If the magnetic monopole solution in Eq.~(\ref{b01}) is chosen as the classical part $b_{\mu}^3$, 
this field cannot couple with the static current $j^{\mu}=(j^0, \vec{0})$.  
Therefore, to study the confinement of color electric charges, we introduce the electric potential 
$\bt_{\mu}$ and its dual potential $\mcb_{\mu}$.  
It is natural to assume the dual relation 
\begin{equation}
     j_{\mu} \leftrightarrow  k_{\mu},\quad \bt_{\mu} \leftrightarrow  \cc_{\mu}, \quad -\mcb_{\mu} \leftrightarrow  \ct_{\mu}.  \lb{301}
\end{equation}
Then Eq.~(\ref{210}) gives the relation 
\begin{align}
 -\epsilon^{\mu\nu\alpha\beta}\ptl_{\alpha}\mcb_{\beta}&=(\ptl\wedge \bt)^{\mu\nu} +\Lambda_e^{\mu\nu}, \lb{302} \\
 \Lambda_e^{\mu\nu}&= -\frac{n^{\mu}}{n^{\rho}\ptl_{\rho}}\ptl_{\sigma}(\ptl\wedge \bt)^{\sigma\nu}
     +\frac{n^{\nu}}{n^{\rho}\ptl_{\rho}}\ptl_{\sigma}(\ptl\wedge \bt)^{\sigma\mu}.  \nonumber
\end{align}
From this equation, we obtain 
\begin{equation}
     (\ptl \wedge \mcb)_{\mu\nu}= \epsilon_{\mu\nu\alpha\beta} 
     \left\{\ptl^{\alpha}\bt^{\beta}-\frac{n^{\alpha}}{n^{\rho}\ptl_{\rho}}\ptl_{\sigma}(\ptl\wedge \bt)^{\sigma\beta}
     \right\}.  \lb{303}
\end{equation}
As in Sect.~2, $\bt_{\mu}$ and $\mcb_{\mu}$ have U(1) symmetries.  
If we choose the gauges
\[
     n_{\mu}\bt^{\mu}=0, \quad  n_{\mu}\mcb^{\mu}=0,  
\]
Eq.~(\ref{303}) leads to 
\begin{equation}
     \mcb^{\nu}=-\epsilon^{\nu\mu\alpha\beta}\frac{n_{\mu}\ptl_{\alpha}}{n^{\rho}\ptl_{\rho}}\bt_{\beta}.  \lb{304}
\end{equation}
Namely, the dual electric potential $\mcb_{\mu}$ has the string singularity.  The term $\Lambda_e^{\mu\nu}$ 
represents the string, which we call the electric string.

     Based on the dual relation in Eq.~(\ref{301}), we can repeat the procedure in Ref.~\cite{hs2}.  So, by 
applying Eq.~(\ref{301}), 
the Lagrangians for $\bt_{\mu}$ and $\mcb_{\mu}$ are obtained from those for $\cc_{\mu}$ and $\ct_{\mu}$.  
However, in this section, we derive them directly.  
To incorporate the electric current $j_{\mu}$, we add 
$\epsilon^{\mu\nu\alpha\beta}(n^{\rho}\ptl_{\rho})^{-1}n_{\alpha}j_{\beta}$ to $(\ptl \wedge \mcb)^{\mu\nu}$.  
\footnote{This is Zwanziger's dual field strength $F^d=(\ptl \wedge B)+(n \cdot \ptl)^{-1}(n \wedge j_e)^d$ in Ref.~\cite{zwa}.}
In addition, taking the London current in superconductivity into account, we add 
$-\epsilon^{\mu\nu\alpha\beta}(n^{\rho}\ptl_{\rho})^{-1}n_{\alpha}m^2\bt_{\beta}/2$ to Eq.(\ref{303}), i.e.,  
\begin{align}
     (\ptl \wedge \mcb)_{\mu\nu}+& \epsilon_{\mu\nu\alpha\beta}\frac{n^{\alpha}}{n^{\rho}\ptl_{\rho}}j^{\beta}
    - \epsilon^{\mu\nu\alpha\beta}\frac{n_{\alpha}}{n^{\rho}\ptl_{\rho}}\frac{m^2}{2}\bt_{\beta} \nonumber \\
     &= \epsilon_{\mu\nu\alpha\beta} \left[\ptl^{\alpha}\bt^{\beta}
    -\frac{n^{\alpha}}{n^{\rho}\ptl_{\rho}}\left\{\ptl_{\sigma}(\ptl\wedge \bt)^{\sigma\beta}
    -j^{\beta}+\frac{m^2}{2}\bt^{\beta}\right\} \right]. \lb{305}
\end{align}
From the left-hand side (LHS) of Eq.(\ref{305}), using Eqs.~(\ref{304}) and (\ref{a02}), we obtain 
\begin{align}
    &-\frac{1}{4} \left[ (\ptl\wedge \mcb)+(n^{\rho}\ptl_{\rho})^{-1}\left\{n\wedge \left(j-\frac{m^2}{2}\bt\right)\right\}^d\right]^2
    \nonumber \\
    &= -\frac{1}{4} \left\{ \ptl\wedge \mcb+(n^{\rho}\ptl_{\rho})^{-1}(n \wedge j)^d\right\}^2+\frac{m^2}{2}\mcb_{\mu}\mcb^{\mu}
    +\frac{m^2}{2}j^{\mu}N_{\mu\nu}\bt^{\nu}- \frac{m^4}{8}\bt^{\mu}N_{\mu\nu}\bt^{\nu}, \lb{306}
\end{align}
where $(A\wedge B)^d_{\mu\nu}=
 \frac{1}{2}\epsilon_{\mu\nu\kappa\lambda} (A\wedge B)^{\kappa\lambda}= 
 \epsilon_{\mu\nu\kappa\lambda}A^{\kappa}B^{\lambda}$.  
In the same way, using the formula 
\[  
    (\ptl \wedge \bt)^{\mu\nu}(n\wedge W)_{\mu\nu}=-2\bt^{\mu}(n^{\rho}\ptl_{\rho})W_{\mu}, 
\]
the right-hand side (RHS) of Eq.(\ref{305}) gives 
\begin{align}
 &-\frac{1}{4} \left[ \left\{ \ptl \wedge\bt -(n^{\rho}\ptl_{\rho})^{-1} n\wedge\left( \ptl(\ptl\wedge \bt)-j+\frac{m^2}{2}\bt 
 \right) \right\}^d \right]^2 \nonumber \\
 &= -\frac{1}{4}(\ptl\wedge \bt)^2 +\frac{m^2}{2}\bt_{\mu}\bt^{\mu}- \bt^{\mu}j_{\mu}
   -\frac{1}{2}\left\{(D_m^{-1})\bt-j \right\}^{\mu}N_{\mu\nu}\left\{(D_m^{-1})\bt-j \right\}^{\nu} \nonumber \\
 &\quad  +\frac{m^2}{2}\left\{(D_m^{-1})\bt-j \right\}^{\mu}N_{\mu\nu}\bt^{\nu}-\frac{m^4}{8}\bt^{\mu}N_{\mu\nu}\bt^{\nu}.  \lb{307}
\end{align}
We note, because of Eq.~(\ref{a01}), the term 
$-\frac{1}{4} \left\{(\ptl \wedge\bt)^d\right\}^2$ gives the kinetic term with the wrong sign \cite{cho}, i.e., 
$\frac{1}{4}(\ptl\wedge \bt)^2$.  The cross term 
\[
 -\frac{1}{2} \left(\ptl \wedge\bt \right)^d
(n^{\rho}\ptl_{\rho})^{-1}\left[ -n \wedge\left\{\ptl(\ptl\wedge \bt)\right\}\right]^d  
\]
changes the sign of this term, and the correct kinetic term is derived \cite{hs2}.    
If we move the last two terms in the LHS (Eq.~(\ref{306})) to the RHS (Eq.~(\ref{307})), we obtain 
\begin{align}
 & -\frac{1}{4} \left\{ \ptl\wedge \mcb + (n^{\rho}\ptl_{\rho})^{-1}(n\wedge j)^d\right\}^2+\frac{m^2}{2}\mcb_{\mu}\mcb^{\mu} \nonumber \\
 &= -\frac{1}{4}(\ptl\wedge \bt)^2 +\frac{m^2}{2}\bt_{\mu}\bt^{\mu}- \bt^{\mu}j_{\mu} + \Omega_j 
   -\frac{1}{2}\left\{(D_m^{-1})\bt-j \right\}^{\mu}N_{\mu\nu}\left\{(D_m^{-1})\bt-j \right\}^{\nu},  \lb{308}
\end{align}
where 
\begin{equation}
       \Omega_j = \frac{m^2}{2}\bt^{\mu}N_{\mu\nu}\left\{(D^{-1}_m)\bt -2j\right\}^{\nu}, \quad
       \{(D^{-1}_m)\bt\}^{\nu}=\ptl_{\lambda}(\ptl \wedge \bt)^{\lambda\nu}+m^2\bt^{\nu}.  \lb{309}
\end{equation}

The LHS of Eq.~(\ref{308}), i.e., 
\begin{equation}
    \cl'_{\mathrm{ecl}}= -\frac{1}{4} \left\{(\ptl\wedge \mcb)^{\mu\nu}
    + \epsilon^{\mu\nu\alpha\beta}\frac{n_{\alpha}}{n^{\sigma}\ptl_{\sigma}}j_{\beta}\right\}
    \left\{(\ptl\wedge \mcb)_{\mu\nu}
    + \epsilon_{\mu\nu\kappa\lambda}\frac{n^{\kappa}}{n^{\rho}\ptl_{\rho}}j^{\lambda}\right\}
    +\frac{m^2}{2}\mcb_{\mu}\mcb^{\mu}  \lb{310}
\end{equation}
is the Lagrangian for $\mcb_{\mu}$.  It gives the equation of motion 
\begin{equation}
     (D_m^{-1})_{\mu\nu}\mcb^{\nu} - \mathcal{J}_{\mu}=0, \quad  
     \mathcal{J}_{\mu}=-\epsilon_{\mu\nu\beta\sigma}\frac{n^{\nu}\ptl^{\beta}}{n^{\rho}\ptl_{\rho}}j^{\sigma}.  \lb{311}
\end{equation}
From the RHS of Eq.~(\ref{308}), we obtain the equation of motion for $\bt_{\mu}$ as 
\begin{equation}
(D_m^{-1})_{\mu\nu}\bt^{\nu} - j_{\mu} +\frac{\delta \Omega_j}{\delta \bt^{\mu}}-
(D_m^{-1})_{\mu\sigma}N^{\sigma\nu}\left\{(D_m^{-1})\bt-j \right\}_{\nu}=0.  \lb{312}
\end{equation}
However this equation is satisfied by the usual equation of motion 
\begin{equation}
     (D_m^{-1})_{\mu\nu}\bt^{\nu} - j_{\mu}=0,  \lb{313}
\end{equation}
because Eq.~(\ref{313}) leads to 
\[ 
  \left.\frac{\delta \Omega_j}{\delta \bt^{\mu}}\right|_{(D_m^{-1})\bt=j}=0.  
\]  
So, using Eq.~(\ref{313}), the RHS of Eq.~(\ref{308}) becomes the following Lagrangian for $\bt_{\mu}$: 
\begin{equation}
    \cl_{\mathrm{ecl}}= -\frac{1}{4} (\ptl\wedge \bt)^2
    +\frac{m^2}{2}\bt_{\mu}\bt^{\mu} - \bt_{\mu}j^{\mu} + \Omega_{j}.  \lb{314}
\end{equation}
We note, if we multiply Eq.~(\ref{313}) by $-\epsilon^{\lambda\kappa\sigma\mu}(n^{\rho}\ptl_{\rho})^{-1}n_{\kappa}\ptl_{\sigma}$, 
Eq.~(\ref{311}) is obtained.

\section{Electric charge confinement}

     Since $\cl_{\mathrm{ecl}}$ is equivalent to $\cl'_{\mathrm{ecl}}$, we consider $\cl_{\mathrm{ecl}}$ first.  
Using the equation of motion $(D_m^{-1})\bt=j$, it becomes 
\begin{equation}
     \cl_{\mathrm{ecl}}= -\frac{1}{2}\bt_{\mu}j^{\mu}-\frac{m^2}{2}\bt^{\mu}N_{\mu\nu}j^{\nu}, \lb{401}
\end{equation}
Substituting $\bt^{\mu}=D_m^{\mu\nu}j_{\nu}$ into Eq.~(\ref{401}), we obtain the electric current-current correlation \cite{suz, mts, sst, sst2}
\begin{equation}
 \cl_{jj}=
 -\frac{1}{2}j_{\mu}\frac{1}{\square + m^2}j^{\mu} 
   -\frac{1}{2}j^{\mu}\frac{m^2}{\square + m^2}\frac{n_{\alpha}n^{\alpha}}{(n^{\rho}\ptl_{\rho})^2}
\left(g_{\mu\nu}-\frac{n_{\mu}n_{\nu}}{n_{\sigma}n^{\sigma}}\right)j^{\nu}.   \lb{402}
\end{equation}
 
     To derive the static potential between the electric charges $Q_e$ and $-Q_e$, 
we insert the static electric current 
\begin{equation}
     j^{\mu}(x)=Q_eg^{\mu 0}\{\delta(\vec{x}-\vec{a})-\delta(\vec{x}-\vec{b})\}.  \lb{403}
\end{equation}
Then the first term in Eq.~(\ref{402}) leads to 
\[
  Q_e^2\int \frac{d^3q}{(2\pi)^3}\frac{1- \cos \vec{q}\cdot \vec{r}}{\vec{q}^2+m^2}, 
\]
where $\vec{r}=\vec{a}-\vec{b}$.  If we write $r=|\vec{r}|$, by removing constants, 
it gives the Yukawa potential 
\[
  V_{\mathrm{eY}}(r)= \frac{-Q_e^2}{4\pi}\frac{e^{-mr}}{r}.  
\]
Next, for the space-like vector $n^{\mu}=(0, \vec{n})$ with $\vec{n}^2=1$, the second term in Eq.~(\ref{402}) 
becomes 
\begin{equation}
     Q_e^2\int \frac{d^3q}{(2\pi)^3}(1- \cos \vec{q}\cdot \vec{r})\frac{m^2}{(\vec{q}^2+m^2)q_n^2},  \lb{404}
\end{equation}
where $q_n=\vec{q}\cdot \vec{n}$.  This expression has the infrared divergence as $q_n \to 0$.  
However, this divergence disappears when $\vec{n}\parallel \vec{r}$ \cite{sst, sst2}.  
\footnote{The infrared behavior of Eq.~(\ref{404}) is discussed in Appendix~C.}
Thus the finite part gives the linear potential 
\begin{equation}
    V_{\mathrm{eL}}(r)=\sigma_e r + O(e^{-mr}), \quad \sigma_e=\frac{Q_e^2m^2}{8\pi}\ln \left(\frac{m^2+m_{\chi}^2}{m^2}\right), 
\lb{405}
\end{equation}
and Eq.~(\ref{402}) leads to the static quark-antiquark potential \cite{suz, mts, sst, sst2}
\begin{equation}
    V_{\mathrm{e}}(r) = V_{\mathrm{eY}}(r)+V_{\mathrm{eL}}(r).   \lb{406}  
\end{equation}

     Instead of the field $\bt^{\mu}$, we can use the dual field $\mcb^{\mu}$.  
If we apply the equation of motion $(D_m^{-1})\mcb=\mathcal{J}$ to $\cl'_{\mathrm{ecl}}$, 
the correlation $\cl_{jj}$ is obtained again.

\section{Origin of the linear potential $V_{\mathrm{eL}}$}

     By multiplying Eq.(\ref{302}) by $n_{\mu}$ and $-j_{\nu}/2$, we obtain 
\begin{equation}
 \frac{1}{2}\left(\epsilon^{\beta\mu\alpha\nu}\frac{n_{\mu}\ptl_{\alpha}}{n^{\rho}\ptl_{\rho}}j_{\nu}\right)\mcb_{\beta}
 =-\frac{1}{2}j_{\nu}\bt^{\nu} +\frac{1}{2}j_{\nu}N^{\nu\mu} \ptl_{\sigma}(\ptl \wedge \bt)^{\sigma}_{\mu},  \lb{501}
\end{equation}
where the gauge condition $n^{\mu}\bt_{\mu}=0$ has been used.  
If we subtract $-j_{\nu}N^{\nu\mu}j_{\mu}/2$, 
Eq.~(\ref{501}) becomes 
\begin{equation}
 -\frac{1}{2}\mathcal{J}^{\beta}\mcb_{\beta}-\frac{1}{2}j_{\nu}N^{\nu\mu}j_{\mu}
 =-\frac{1}{2}j_{\nu}\bt^{\nu}+ \frac{1}{2}j_{\nu}N^{\nu\mu} \ptl_{\sigma}(\ptl \wedge \bt)^{\sigma}_{\mu}
 -\frac{1}{2}j_{\nu}N^{\nu\mu}j_{\mu}.  \lb{502}
\end{equation}
where $\mathcal{J}^{\beta}$ is defined in Eq.~(\ref{311}).  
The equation of motion 
$\ptl_{\sigma}(\ptl \wedge \bt)^{\sigma}_{\mu}+m^2\bt_{\mu}-j_{\mu}=0$ makes the RHS of Eq.~(\ref{502}) 
\begin{equation}
  -\frac{1}{2}j_{\mu}\bt^{\mu}- \frac{m^2}{2}j_{\nu}N^{\nu\mu} \bt_{\mu}.  \lb{503}
\end{equation}
Since the current conservation $\ptl_{\mu}j^{\mu}=0$ leads to 
$\bt_{\mu}=(\Box+m^2)^{-1}j_{\mu}$, Eq.~(\ref{503}) becomes 
\begin{equation}
     -\frac{1}{2}j^{\mu}\frac{1}{\Box +m^2} j_{\mu} -\frac{m^2}{2}j^{\mu}\frac{1}{\Box +m^2}N_{\mu\nu}j^{\nu}.  \lb{504}
\end{equation}
The first term and the second term yield the Yukawa potential and the linear potential, respectively.

     The factor 
$\frac{1}{2}j_{\nu}N^{\nu\mu} \ptl_{\sigma}(\ptl \wedge \bt)^{\sigma}_{\mu}$ comes from the electric string 
$\Lambda^{\mu\nu}_{e}$ in Eq.~(\ref{302}).  This factor becomes $- \frac{m^2}{2}j_{\nu}N^{\nu\mu} \bt_{\mu}$ only when $m\neq 0$.  
Therefore there are two causes of the linear potential.  One is the electric string 
and the other is the mass for the electric potential.

\section{Classical configuration for confinement}

     In Eq.~(\ref{201}), we can choose any classical solution as $b_{\mu}^3$.  First we choose 
$\bt_{\mu}$ as $b_{\mu}^3$.  The coupling with the electric current is supposed to be $-j^{\mu}\bt_{\mu}$.  
Then the classical part of Eq.~(\ref{201}) gives 
\begin{equation}
-\frac{1}{4}(\ptl \wedge \bt)^2 +\frac{m^2}{2}\bt_{\mu}\bt^{\mu} -j^{\mu}\bt_{\mu}, \lb{601}
\end{equation}
and the equation of motion $(D_m^{-1})_{\mu\nu}\bt^{\nu}=j_{\mu}$ is satisfied.  

    Next we consider another solution $\mathfrak{B}_{\mu}(\bt,n)$, which contains $\bt_{\mu}$ and $n^{\mu}$.  
To couple with $\mathfrak{B}_{\mu}$, the current $j^{\mu}$ may be modified.  
This modified current, which depends on $j^{\mu}$ and $n^{\mu}$, is denoted by $\mathfrak{J}^{\mu}(j,n)$.  
Then, by setting $b_{\mu}^3=\mathfrak{B}_{\mu}$, Eq.~(\ref{201}) gives 
\begin{equation}
     -\frac{1}{4}(\ptl \wedge \mathfrak{B})^2 +\frac{m^2}{2}\mathfrak{B}^{\mu}\mathfrak{B}_{\mu} -\mathfrak{J}^{\mu}\mathfrak{B}_{\mu}. \lb{602}
\end{equation}
Now we assume that Eq.~(\ref{602}) is rewritten as 
\begin{equation}
-\frac{1}{4}(\ptl \wedge \bt)^2 +\frac{m^2}{2}\bt_{\mu}\bt^{\mu} -j^{\mu}\bt_{\mu} +  \Delta \cl(\bt,n,j).  \lb{603}
\end{equation}
If $\Delta \cl(\bt,n,j)$ satisfies 
\[
  \left.\frac{\delta \Delta \cl(\bt,n,j)}{\delta \bt^{\mu}} \right|_{(D_m^{-1})\bt=j} =0,  
\]
Eq.~(\ref{603}) yields the same equation of motion $(D_m^{-1})_{\mu\nu}\bt^{\nu}=j_{\mu}$.  
Thus, although Eqs.~(\ref{601}) and (\ref{603}) produce the same equation of motion, because of the 
term $\Delta \cl(\bt,n,j)$, some additional effects may exist.  

     This is the situation studied in the sections 3 and 4.  We set $\mathfrak{B}_{\mu}=\mcb_{\mu}$ and $\mathfrak{J}^{\mu}=\mathcal{J}^{\mu}$  
defined in Sect.~3.  Then we find 
\[
  \Delta \cl(\bt,n,j)=\Omega_j 
   -\frac{1}{2}\left\{(D_m^{-1})\bt-j \right\}^{\mu}N_{\mu\nu}\left\{(D_m^{-1})\bt-j \right\}^{\nu}
   +\frac{1}{2}j^{\mu}N_{\mu\nu} j^{\nu},  
\]
and the term $\Omega_j $ yields the confining potential.  

     Thus we can conclude that the classical configuration which yields the quark confinement is the 
monopole solution of the dual gauge field $\mcb_{\mu}$.

\section{Comparison with the dual Ginzburg-Landau model}

     Zwanziger considered a local Lagrangian with two gauge fields $A^{\mu}$ and $B^{\mu}$ \cite{zwa}.  If 
$B^{\mu}$ is integrated out, it gives the nonlocal Lagrangian \cite{bs}
\begin{equation}
   \cl(A)=-\frac{1}{4}(F)^2 - j_{\mu}A^{\mu}, \quad F=\ptl \wedge A  -(n^{\rho}\ptl_{\rho})^{-1}(n\wedge k)^d. \lb{701}
\end{equation}
In the same way, we can derive the equivalent Lagrangian 
\begin{equation}
   \cl(B)=-\frac{1}{4}(F^d)^2 - k_{\mu}B^{\mu}, \quad F^d=\ptl \wedge B + (n^{\rho}\ptl_{\rho})^{-1}(n\wedge j)^d. \lb{702}
\end{equation}

     To study the quark confinement, $\cl(B)$ is often used.  In the dual Ginzburg-Landau model of dual superconductor, 
introducing the monopole field $\chi$, 
replace the term $- k_{\mu}B^{\mu}$ with the covariant derivative of $\chi$ \cite{bss}.  This part contains the term $|B_{\mu}\chi|^2$.  
Adding an appropriate potential $V(\chi)$, the VEV $\langle \chi \rangle$ appears, and $B_{\mu}$ becomes massive.  
Then the interaction $(\ptl \wedge B)(n^{\rho}\ptl_{\rho})^{-1}(n\wedge j)^d$ in $(F^d)^2$ produces the linear potential \cite{mts}.  

     In this case, 
we can identify $A^{\mu}$ and $B^{\mu}$ with $\ct^{\mu}=(\ct^0, \vec{\ct})$ and $\cc^{\mu}=(\cc^0, \vec{\cc})$, 
respectively.  Let us consider the current $j^{\mu}=(j^0,\vec{0})$.  
Since $j^{0}$ couples with not $\cc^0$ but $\vec{\cc}$, the component $\vec{\cc}$ is indispensable 
to produce the linear potential $V_{\mathrm{eL}}$.  The coupling between the magnetic current $k^{\mu}=(k^0,\vec{k})$ and 
$\cc^{\mu}$ is $k^{\mu}\cc_{\mu}$, the space component $\vec{k}$ is also necessary.  
Furthermore, to make $\vec{\cc}$ massive, some additional mechanism like the introduction of $\chi$ and $V(\chi)$ 
is inevitable.  

     In the present approach, we can identify $A^{\mu}$ and $B^{\mu}$ with 
$\bt^{\mu}=(\bt^0, \vec{0})$ and $\mcb^{\mu}=(0, \vec{\mcb})$.  
Since the ghost condensation and the VEV $\langle A_{\mu}^+A^{-\mu}\rangle$ 
produce the mass for any classical solution, the magnetic current $k^{\mu}$ and additional fields like $\chi$ are unnecessary 
to yield $V_{\mathrm{eL}}$.  The Lagrangian $\cl(B)$ holds by adding the mass term $m^2B^2/2$.  
However, as we showed in Sect.~3, in addition to the mass term $m^2A^2/2$, 
the term $\Omega_j$ in Eq.~(\ref{309}) should be added to the Lagrangian $\cl(A)$.  

     We make a comment.  For the currents $j^{\mu}=(j^0, \vec{0})$ and $k^{\mu}=(k^0, \vec{0})$, 
it is possible to set $A^{\mu}=(\bt^0, \vec{\ct})$ and $B^{\mu}=(\cc^0, \vec{\mcb})$.  
When $j^0=0$, $\bt^0$ and $\vec{\mcb}$ vanish, and magnetic charges are confined \cite{hs2}.  
Likewise, if $k^0=0$, $\vec{\ct}$ and $\cc^0$ vanish, and electric charges are confined.

\section{Summary and comments}

     In the previous papers \cite{hs1, hs2, hs3}, we studied the SU(2) gauge theory in the nonlinear gauge 
of the Curci-Ferrari type.  It was shown that, because of the ghost condensation $\vp_0\neq 0$, 
the SU(2) gauge theory breaks down to the U(1) theory in the low-energy region \cite{hs1}.  
In Ref.~\cite{hs}, we found that, although the quantum U(1) gauge field $a_{\mu}^3$ 
is massless, the classical part $b_{\mu}^3$ acquires the mass $m=\sqrt{g^3\vp_0/(32\pi)}$ through the VEV 
$\langle A_{\mu}^+A^{-\mu}\rangle$.  
Then, in Ref.~\cite{hs2}, we considered the magnetic potential $\ct_{\mu}$ as the classical part $b_{\mu}^3$.  
It was shown that the magnetic charges $Q_m$ and $-Q_m$ are confined by the linear potential.  We also showed that the 
linear potential is derived by using the dual magnetic potential $\cc_{\mu}$ consistently.  

     In this paper, we considered the electric potential $\bt_{\mu}$ and the dual electric potential $\mcb_{\mu}$ 
as the classical part $b_{\mu}^3$.  
The dual relation between $\bt_{\mu}$ and $\mcb_{\mu}$ requires the expression (\ref{302}), which contains 
the string term $\Lambda_{e}^{\mu\nu}$.  In fact, for a point 
color electric charge, $\bt_{\mu}$ is the usual Coulomb potential, $\mcb_{\mu}$ is the monopole-type potential, and 
$\Lambda_{e}^{\mu\nu}$ is the electric Dirac string.  
Using Eq.~(\ref{302}), we derived the Lagrangians 
$\cl_{\mathrm{ecl}}$ with $\bt_{\mu}$ and $\cl'_{\mathrm{ecl}}$ with $\mcb_{\mu}$, and showed the relation 
$\cl_{\mathrm{ecl}}=\cl'_{\mathrm{ecl}}$.  We note, by applying the duality (\ref{301}), 
the Lagrangians $\cl'_{\mathrm{mAbel}}$ and $\cl_{\mathrm{mAbel}}$ in Sect.~2 lead to the Lagrangians $\cl_{\mathrm{ecl}}$ and 
$\cl'_{\mathrm{ecl}}$, respectively.  

     From the Lagrangian $\cl_{\mathrm{ecl}}$, the linear potential $V_{\mathrm{eL}}$ 
between the color electric charges $Q_e$ and $-Q_e$ is obtained.  The operator $1/(n^{\rho}\ptl_{\rho})$, which 
yields the unphysical string, and the mass $m$ are necessary to give $V_{\mathrm{eL}}$.  
The Lagrangian $\cl_{\mathrm{ecl}}$ contains the 
term $\Omega_j$ in Eq.~(\ref{309}).  This term, which is the origin of the linear potential, comes from 
the electric string $\Lambda^{\mu\nu}_{e}$ and the mass term for $\bt_{\mu}$.  We can also use the Lagrangian 
$\cl'_{\mathrm{ecl}}$ to derive $V_{\mathrm{eL}}$.  For a point color electric charge, $\mcb_{\mu}$ is the 
monopole-type solution.  So we can say that 
the classical configuration which yields the quark confinement is the monopole solution of the dual gauge field $\mcb_{\mu}$.  

     In the dual Ginzburg-Landau model of dual superconductor, the operator $1/(n^{\rho}\ptl_{\rho})$ exists as well.  
However, there are two different points.  One is the fields that contribute, and the other is the mechanism to 
produce the mass $m$.  In the dual superconductor model, the field (dual field) is $\ct_{\mu}$ ($\cc_{\mu}$), and 
the mechanism is the monopole condensation.  In the present approach, the field (dual field) is $\bt_{\mu}$ ($\mcb_{\mu}$), 
and the mechanism is the condensation $\langle A_{\mu}^+A^{-\mu}\rangle$ subsequent to the ghost condensation $\vp_0\neq 0$.

     We make some comments.

(1).  Below the scale $\Lambda_{\mathrm{QCD}}$, the ghost $c^A$ and the antighost $\bar{c}^A$ make a bound state $igf^{ABC}\bar{c}^B c^C$ \cite{hs4}, 
and the ghost condensate $\vp_0$ appears.  This is the origin of the mass $m$.  
This condensation happens in the non-Abelian gauge theory.  Without $m$, the term $\Omega_j$ vanishes, and the Lagrangian 
$\cl_{\mathrm{ecl}}$ in Eq.~(\ref{314}) reduces to the usual U(1) Lagrangian.  Thus the confinement by the 
mechanism presented here does not happen in QED. 

(2).  The operator $1/(n^{\rho}\ptl_{\rho})$ yields the string singularity.  This singularity should not be 
detected.  However, as we stated in Sect.~3 and stressed in Ref.~\cite{hs2}, the string term $\Lambda_{e}^{\mu\nu}$ is important 
to yield the correct kinetic term $-(\ptl\wedge \bt)^2/4$.  In addition, the effect of the string exists energetically.  
The energy of the string is proportional to its length \cite{cgpz}.  
This is the infrared divergence $\propto 1/\varepsilon$ in Appendix~C.  To get a finite energy, the color electric 
charges $Q_e$ and $-Q_e$ must be on the line determined by $\vec{n}$, where $n^{\mu}=(0,\vec{n})$.  Then the energy becomes finite, 
and is proportional to the distance $r=|\vec{r}|$ between them.  

(3).  Physical quantities should not depend on $n^{\mu}$.  For example, the equation of motion for $\bt_{\mu}$ 
presented in Eq.~(\ref{312}) contains $n^{\mu}$.  However, it reduces to the usual equation of motion 
$(D_m^{-1})_{\mu\nu}\bt^{\nu}=j_{\mu}$.  
The next example is the linear potential.  The positions of the charges $Q_e$ and $-Q_e$ must satisfy 
$\vec{r}\parallel \vec{n}$ energetically.  However, as $\vec{n}$ can be chosen in an arbitrary direction, 
we can put $Q_e$ and $-Q_e$ in arbitrary positions, and the potential 
$V_{\mathrm{eL}}(r)$ is independent of $n^{\mu}$.

\appendix

\section{Notations and some relations}

     We employ the metric $g_{\mu\nu}=\mathrm{diag}(1,-1,-1,-1)$.  The antisymmetric pseudotensor 
$\epsilon^{\mu\nu\rho\sigma}$ defined by $\epsilon^{0123}=1$ satisfies the formulas 
\begin{equation}
    \epsilon^{\alpha\beta\rho\sigma}\epsilon_{\alpha\lambda\mu\nu}=-\left|
\begin{array}{ccc}
 \delta^{\beta}_{\lambda}&\delta^{\beta}_{\mu} & \delta^{\beta}_{\nu} \\
 \delta^{\rho}_{\lambda}&\delta^{\rho}_{\mu} & \delta^{\rho}_{\nu} \\
\delta^{\sigma}_{\lambda} & \delta^{\sigma}_{\mu}& \delta^{\sigma}_{\nu}
\end{array}
\right|,\quad
     \epsilon^{\alpha\beta\rho\sigma}\epsilon_{\rho\sigma\mu\nu}=-2(\delta^{\alpha}_{\mu}\delta^{\beta}_{\nu}
     -\delta^{\alpha}_{\nu}\delta^{\beta}_{\mu}).   \lb{a01} 
\end{equation}
From Eq.~(\ref{a01}), the following relations are obtained: 
\begin{align}
     \epsilon_{\mu\nu\alpha\beta}\frac{n^{\alpha}J^{\beta}}{n^{\rho}\ptl_{\rho}}
     \epsilon^{\mu\nu\kappa\lambda}\frac{n_{\kappa}K_{\lambda}}{n^{\sigma}\ptl_{\sigma}}
     =&
2J^{\mu}\frac{n_{\beta}n^{\beta}}{(n^{\rho}\ptl_{\rho})^2}\left(g_{\mu\nu}-\frac{n_{\mu}n_{\nu}}{n_{\sigma}n^{\sigma}}\right)K^{\nu}
=2J^{\mu}N_{\mu\nu}K^{\nu},  \lb{a02} \\
    \epsilon_{\mu\nu\alpha\beta}\frac{n^{\nu}\ptl^{\alpha}J^{\beta}}{n^{\rho}\ptl_{\rho}}
\epsilon^{\mu\eta\kappa\lambda}\frac{n_{\eta}\ptl_{\kappa}K_{\lambda}}{n^{\sigma}\ptl_{\sigma}}
 =& J^{\mu}g_{\mu\nu}K^{\nu} - J^{\mu}\Box N_{\mu\nu}K^{\nu}  \nonumber \\
  &+ J^{\mu}\frac{1}{(n^{\rho}\ptl_{\rho})^2} \left( n_{\sigma}n^{\sigma}\ptl_{\mu}\ptl_{\nu}  
  - n^{\sigma}\ptl_{\sigma}\ptl_{\mu}n_{\nu}-n^{\sigma}\ptl_{\sigma}n_{\mu}\ptl_{\nu} \right) K^{\nu},  \lb{a03} 
\end{align}
where 
\[
N_{\mu\nu}=\frac{n^{\alpha}n_{\alpha}}{(n^{\rho}\ptl_{\rho})^2}
\left(g_{\mu\nu}-\frac{n_{\mu}n_{\nu}}{n^{\beta}n_{\beta}}\right).  
\]
     For simplicity, we use the notations 
\begin{align}
 & (\ptl\wedge \cc)^{\mu\nu}=\ptl^{\mu}\cc^{\nu}-\ptl^{\nu}\cc^{\mu},\quad 
     \ptl_{\sigma}(\ptl\wedge \cc)^{\sigma\nu}=\Box \cc^{\nu}-\ptl^{\nu}\ptl_{\sigma}\cc^{\sigma},  \lb{a04} \\
 & (A\wedge B)^d_{\mu\nu}= \frac{1}{2}\epsilon_{\mu\nu\kappa\lambda} (A\wedge B)^{\kappa\lambda}= \epsilon_{\mu\nu\kappa\lambda}A^{\kappa}B^{\lambda}, \nonumber
\end{align}
and, for an antisymmetric tensor $H_{\mu\nu}$, 
\[
 H^2 = H_{\mu\nu}H^{\mu\nu}.  
\]

\section{Monopole solutions and dual solutions}

     For a magnetic charge and an electric charge, we present monopole solutions and dual solutions 
in the massless case and the massive case.  

\subsection{Magnetic potential and its dual potential}

     In the massless case, we choose the magnetic potential, which describes a magnetic monopole, as 
\begin{equation}
     \ct_{\mu}=\frac{N}{g}\frac{z-r}{r\rho^2}(0,-y,x,0),   \lb{b01}
\end{equation}
where $N$ is an integer, and $\rho=\sqrt{x^2+y^2}$.  This field satisfies the equation 
\begin{align*}
    &\ptl_{\mu}(\ptl \wedge \ct)^{\mu\nu} -
\epsilon^{\nu\alpha\mu\beta}\frac{n_{\alpha}\ptl_{\mu}}{n^{\rho}\ptl_{\rho}}k_{\beta}=0,  \\
    & n^{\alpha}=\delta^{\alpha}_3,\quad k^{\beta}=\frac{4\pi N}{g} \delta(x)\delta(y)\delta(z)\delta^{\beta}_0 . 
\end{align*}
The corresponding dual magnetic potential $\cc^{\mu}$ and its equation of motion are 
\begin{equation}
     \cc^{\mu}=\frac{N}{g}\frac{1}{r}\delta^{\mu}_{0}, \quad 
     \ptl_{\mu}(\ptl \wedge \cc)^{\mu\nu}-k^{\nu}=0.  \lb{b02}
\end{equation}

     From Eq.~(\ref{b02}), the dual field strength $\hh^{\mu\nu}$ in Eq.~(\ref{210}) becomes  
\[
      \hh^{0j}=-\ptl^j\cc^0=-\frac{N}{g}\frac{x^j}{r^3}\ (j=1,2),\quad 
       \hh^{03}=-\frac{N}{g}\frac{x^3}{r^3}+\Lambda_m^{03}.  
\]
We follow Zwanziger's definition \cite{zwa} 
\[
     \frac{1}{\ptl_z}f(x,y,z)=a\int_0^{\infty} f(x,y,z-s)ds -(1-a)\int_0^{\infty}f(x,y,z+s)ds,  
\]
and, to put the Dirac string on the negative $z$-axis, set $a=0$.   
This choice gives 
\begin{equation}
      \frac{1}{\ptl_z}\delta(z-b)=-\theta(b-z),  \lb{b03}
\end{equation}
and we find 
\[  \Lambda_m^{03}=\frac{1}{\ptl_z}(\Box \cc^0)= -\frac{4\pi N}{g}\theta(-z)\delta(x)\delta(y).  \]
Namely, $\Lambda_m^{\mu\nu}$ represents the Dirac string part.  

     In the massive case, Eq.~(\ref{b01}) changes to 
\[
     \ct_{\mu}=\frac{N}{g}\frac{z-r}{r\rho^2}e^{-mr}(0,-y,x,0),   
\]
and it fulfills the equation 
\[
    \ptl_{\mu}(\ptl \wedge \ct)^{\mu\nu} + m^2\ct^{\nu} -
\epsilon^{\nu\alpha\mu\beta}\frac{n_{\alpha}\ptl_{\mu}}{n^{\rho}\ptl_{\rho}}k_{\beta}=0. 
\]
The dual magnetic potential and its equation of motion change from Eq.~(\ref{b02}) to 
\[
     \cc^{\mu}=\frac{N}{g}\frac{e^{-mr}}{r}\delta^{\mu}_{0}, \quad 
     \ptl_{\mu}(\ptl \wedge \cc)^{\mu\nu} + m^2\cc^{\nu} -k^{\nu}=0.   
\]

\subsection{Electric potential and its dual potential}

     Next we consider the color electric current 
\[    j^{\beta}= g \delta(x)\delta(y)\delta(z)\delta^{\beta}_0, \]  
and apply the dual relation (\ref{301}).  In the massless case, the electric potential 
$\bt^{\mu}$ and its dual potential $\mcb_{\mu}$ given by 
\begin{equation}
     \bt^{\mu}=\frac{g}{4\pi}\frac{1}{r}\delta^{\mu}_{0}, \quad 
     \mcb_{\mu}=-\frac{g}{4\pi}\frac{z-r}{r\rho^2}(0,-y,x,0).  \lb{b04}
\end{equation}
satisfy the equations 
\begin{equation}
     \ptl_{\mu}(\ptl \wedge \bt)^{\mu\nu}-j^{\nu}=0  \lb{b05}
\end{equation}
and 
\begin{equation}
      \ptl_{\mu}(\ptl \wedge \mcb)^{\mu\nu} +
\epsilon^{\nu\alpha\mu\beta}\frac{n_{\alpha}\ptl_{\mu}}{n^{\rho}\ptl_{\rho}}j_{\beta}=0.    \lb{b06}
\end{equation}

     Using $\bt^{\mu}$ in Eq.~(\ref{b04}), we find the RHS of Eq.~(\ref{302}), i.e., $(\ptl \wedge \bt)^{0j} +\Lambda_{e}^{0j}$ becomes 
\[ (\ptl \wedge \bt)^{0j}=-\frac{g}{4\pi}\frac{x^j}{r^3},\quad \Lambda_e^{0j}=-\frac{g}{4\pi}\theta(-z)\delta(x)\delta(y)\delta^{3j}.  \]
Namely, $\Lambda_e^{\mu\nu}$ represents the string part.  

     In the massive case, the potential $\bt^{\mu}$ and the dual potential $\mcb_{\mu}$ in Eq.~(\ref{b04}) change to 
\begin{equation}
     \bt^{\mu}=\frac{g}{4\pi}\frac{e^{-mr}}{r}\delta^{\mu}_{0}, \quad 
      \mcb_{\mu}=-\frac{g}{4\pi}\frac{z-r}{r\rho^2}e^{-mr}(0,-y,x,0).  \lb{b07}
\end{equation}
Instead of Eqs.~(\ref{b05}) and (\ref{b06}), they fulfill the equations 
\[
     \ptl_{\mu}(\ptl \wedge \bt)^{\mu\nu} + m^2\bt^{\nu} -j^{\nu}=0  
\]
and 
\[
    \ptl_{\mu}(\ptl \wedge \mcb)^{\mu\nu} + m^2\mcb^{\nu} +
\epsilon^{\nu\alpha\mu\beta}\frac{n_{\alpha}\ptl_{\mu}}{n^{\rho}\ptl_{\rho}}j_{\beta}=0,  
\]
respectively.

\section{Calculation of Eq.~(\ref{404})}

     First, we calculate the integral 
\[ \int_C \frac{e^{izr_n}}{(z^2+\omega^2)z^2}dz  \]
along the path $C$ in Fig.~C1.  In the limit $R\to \infty$ and $\varepsilon \to +0$, this integral 
gives 

\begin{figure}
\begin{center}
\includegraphics{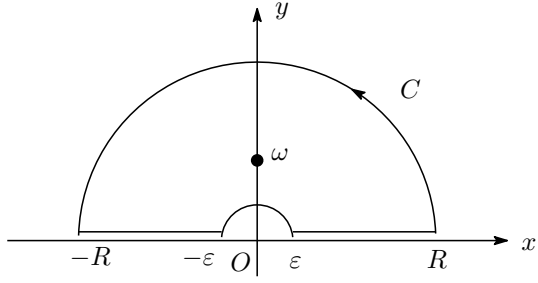}
\caption{The path $C$ on the complex plane.}
\label{figC1}
\end{center}
\end{figure}

\[ I_1(r_n, \omega)= \mathcal{P} \int_{-\infty}^{\infty} \frac{\cos (q_n r_n)}{(q_n^2+\omega^2)q_n^2}dq_n
  =\frac{2}{\varepsilon \omega^2} - \frac{\pi r_n}{\omega^2}- \frac{\pi}{\omega^3}e^{-\omega r_n}, \]
where we set $r_n \geq 0$, and $ \mathcal{P}$ means the Cauchy principal value.  
Using $I_1$ with $\omega=\sqrt{q_T^2+m^2}$, Eq.~(\ref{404}) becomes  
\begin{equation}
   V_1(r_n,r_T)= \frac{Q_e^2m^2}{(2\pi)^3}\int d^2q_T \left\{ I_1(0,\omega)-I_1(r_n,\omega)\cos (\vec{q}_T\cdot\vec{r}_T) \right\},  \lb{c01}
\end{equation}
where $\vec{q}_T\cdot \vec{n}=0$ and $\vec{r}_T \cdot \vec{n}=0$.  We note, when $r_T=|\vec{r}_T|=0$, the infrared divergences 
$2/(\varepsilon \omega^2)$ in $I_1(0,\omega)$ and $I_1(r_n,\omega)$ cancel out.  
Using the Bessel function $J_0(ax)$, we define $K_0(ya,m_{\chi})$ as 
\[   K_0(ya,m_{\chi})=\int_0^{m_{\chi}} dx \frac{x}{x^2+y^2}J_0(ax), \quad J_0(ax)=\frac{1}{2\pi}\int_{0}^{2\pi} d\phi e^{-iax \cos \phi}.  
\]
This function satisfies 
\[ K_0(ya)=\lim_{m_{\chi}\to \infty} K_0(ya,m_{\chi}),\quad \lim_{a\to +0}K_0(ya,m_{\chi}) = \frac{1}{2}\ln \left(\frac{m_{\chi}^2+m^2}{m^2}\right),  \]  
where $K_0(ya)$ is the modified Bessel function.  Then Eq.~(\ref{c01}) becomes 
\begin{align}
 V_1(r_n,r_T)= & \frac{Q_e^2m^2}{4\pi^2\varepsilon}\left\{\ln\left(\frac{m_{\chi}^2+m^2}{m^2}\right)-2K_0(mr_T,m_{\chi})\right\} \nonumber  \\
     & +\frac{Q_e^2m^2}{4\pi}K_0(mr_T,m_{\chi})r_n + I_2(r_n) + C_1,   \lb{c02}
\end{align}
where $m_{\chi}$ is the ultraviolet cutoff for $\vec{q}_T$, and 
\[ I_2(r_n)=\frac{Q_e^2m^2}{8\pi^2}\int d^2q_T \frac{e^{-\omega r_n}}{\omega^3}\cos(\vec{q}_T\cdot\vec{r}_T),\quad
  C_1=-\frac{Q_e^2m}{4\pi}\left(1-\frac{m}{\sqrt{m_{\chi}^2+m^2}}\right).  
\]
When $r_T\neq 0$, as $\ln [(m_{\chi}^2+m^2)/m^2]>2K_0(mr_T,m_{\chi})$, the infrared divergence $1/\varepsilon$ 
exists, and we obtain $\lim_{\varepsilon \to +0}V_1(r_n,r_T)=\infty$.   
On the other hand, if $r_T = 0$, this divergence disappears.   
Since $I_2(r_n)$ satisfies \cite{mts}
\[ |I_2(r_n)|\leq \frac{Q_e^2m}{4\pi}e^{-mr_n},  \]
neglecting $I_2$ and the constant $C_1$, $V_1(r, 0)$ becomes $V_{\mathrm{eL}}(r)$ in 
Eq.~(\ref{405}).  

     We note the damping behaviors $K_0(mr_T)$ and $e^{-mr_n}$ come from the propagator $1/(\Box +m^2)$ in 
$j^{\mu}N_{\mu\nu}/(\Box +m^2)j^{\nu}$, and the behavior $\propto r_n$ is from the operator $N_{\mu\nu}$.  

\begin{figure}
\begin{center}
\includegraphics{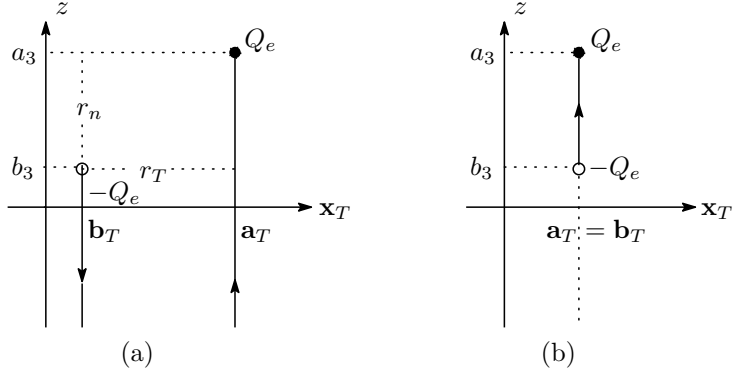}
\caption{The relation between the length $r_T$ and the potential $V_1(r_n,r_T)$ in Eq.~(\ref{c02}).  
The cases with $r_T\neq 0$ and $r_T=0$ are depicted in (a) and (b), respectively.}
\label{figC2}
\end{center}
\end{figure}

     To see the meaning of the above infrared divergence, 
we choose $\vec{n}=(0,0,1)$, $\vec{a}=(\vec{a}_T,a_3)$, and $\vec{b}=(\vec{b}_T,b_3)$.  
Using Eq.~(\ref{b03}), the current (\ref{403}) gives 
\[ \frac{1}{\ptl_z}j^{\mu}(x)=-Q_eg^{\mu 0}\left\{\delta(\vec{x}_T-\vec{a}_T)\theta(a_3-z)
  -\delta(\vec{x}_T-\vec{b}_T)\theta(b_3-z) \right\}.  \]
When $\vec{a}_T\neq \vec{b}_T$, the strings from $Q_e$ and $-Q_e$ with infinite length exist.  This is the 
origin of the infrared divergence.  However, when $\vec{a}_T= \vec{b}_T$, the strings 
with infinite length disappear, and there remains the string with the length $|\vec{a}-\vec{b}|$.  
This situation is depicted in Fig.~C2.

\section{On the ghost condensation in Minkowski space}

     From the ghost determinant $\det(\Box + g\vp_0 \times )$, we obtain the potential 
\begin{align}
 iV_{\mathrm{gh}}(v) =& i\int \frac{d^4p}{(2\pi)^{4}} \ln [(-p^2-i \epsilon)^2 +v^2] \nonumber \\
 =& i\int \frac{d^4p}{(2\pi)^{4}} \ln [(-p^2-i \epsilon -iv)(-p^2-i\epsilon  +iv)],  \lb{d01}
\end{align}
where $v=g\vp_0$.  When we calculate it, there are two cases, i.e., $\epsilon > v$ and $\epsilon < v$.  

     In the case of $\epsilon > v$, since $\mathrm{Im}(p^2+i(\epsilon \pm v))>0$, 
the usual Wick rotation is applicable to the $p^0$-integral.  Using the dimensional regularization, 
we obtain 
\begin{align}
     iV_{\mathrm{gh}}(v)=&-\frac{1}{(4\pi)^2}\left(\frac{1}{\varepsilon}-\gamma +\ln 4\pi +\frac{3}{2}\right)
     (v^2+ \epsilon^2)  \nonumber \\
     & +\frac{1}{2(4\pi)^2}(v^2+ \epsilon^2)\ln (v^2-\epsilon^2) 
     -\frac{1}{(4\pi)^2} \epsilon v \ln\left(\frac{\epsilon -v}{\epsilon + v}\right).  \lb{d02}
\end{align}
This potential is analytically continued to the region $\epsilon < v$, and we can set $\epsilon =0$.  
Then $iV_{\mathrm{gh}}(v)$ becomes the real potential 
\begin{equation}
-\frac{1}{(4\pi)^2}\left\{\left(\frac{1}{\varepsilon}-\gamma +\ln 4\pi +\frac{3}{2}\right)
     v^2 -\frac{1}{2}v^2\ln v^2 \right\},  \lb{d03}
\end{equation}
This potential coincides with the one obtained in Euclidean space, and 
leads to the condensation $v \neq 0$ \cite{hs3}.  

     In the case of $\epsilon < v$, we can apply the Wick rotation for $p^2+i(\epsilon + v)$.  However, 
for $p^2+i(\epsilon - v)$, the rotation must be done in the counter direction.  Thus we obtain 
\begin{align}
     iV_{\mathrm{gh}}(v)=&-\frac{2}{(4\pi)^2}\left(\frac{1}{\varepsilon}-\gamma +\ln 4\pi +\frac{3}{2}\right)
     \epsilon v  \nonumber \\
     & +\frac{1}{2(4\pi)^2}(v^2+ \epsilon^2)\ln\left(\frac{\epsilon +v}{\epsilon - v}\right)
     + \frac{1}{(4\pi)^2}\epsilon v \ln (v^2-\epsilon^2) ,   \lb{d04}
\end{align}
which is complex in the limit $\epsilon \to 0$.  

\begin{figure}
\begin{center}
\includegraphics{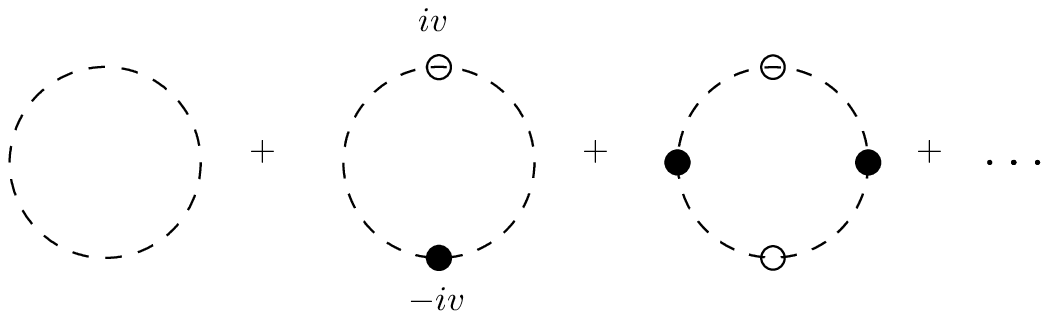}
\caption{The one-loop ghost diagrams.  The dashed line is the ghost propagator $\langle c^{\mp}\cb^{\pm}\rangle$, and 
the blobs represent $\pm iv$.}
\label{figD1}
\end{center}
\end{figure}

    The one-loop diagrams in Fig.D1 lead to the series 
\begin{equation}
    \int \frac{d^4p}{(2\pi)^{4}i} \ln(-p^2 -i\epsilon)^2 + 
     \sum_{n=1}^{\infty} \frac{-1}{n}\int \frac{d^4p}{(2\pi)^{4}i} \left\{-\frac{v^2}{(-p^2-i\epsilon)^2}\right\}^n.  \lb{d05}
\end{equation}
The series 
\[
\sum_{n=1}^{\infty} \frac{-1}{n} \left\{-\frac{v^2}{(-p^2-i\epsilon)^2}\right\}^n  
\]
converges under the condition $\epsilon > v$ for an arbitrary value of $p^2$, and Eq.~(\ref{d05}) becomes 
Eq.~(\ref{d01}).  When we calculate Eq.~(\ref{d05}), since $\mathrm{Im}(p^2+i\epsilon)>0$, 
the usual Wick rotation is used.  Thus Eq.~(\ref{d05}) gives Eq.~(\ref{d03}), and 
the condensate $v\neq 0$ appears \cite{hs2}.  

     In Ref.~\cite{hs3}, without $\epsilon$, we calculated Eq.~(\ref{d01}) directly.  This implies that 
Eq.~(\ref{d04}) with $\epsilon =0$ was obtained.  However this result is not equivalent to Eq.~(\ref{d05}).  
The contribution of the residues at $p^0=\pm \sqrt{\vec{p}^2+i\epsilon}$ is missing.

\end{document}